\documentclass[12pt,a4paper]{article}
\pdfoutput=1
\usepackage{jheppub}

\newcommand{\conj}[1]{\overline{#1}}         
\newcommand{\til}{\widetilde}                
\renewcommand{\hat}{\widehat}                  
\newcommand{\pint}{\makebox[0pt][l]{\hspace{3pt}$-$}\int} 
\newcommand{\EQ}[1]{\begin{align}\begin{split} #1 
\end{split}\end{align}}
\newcommand{\eq}[1]{\begin{equation} #1 
\end{equation}}

\title{Scattering of Giant Holes}
\author[]{Nick Dorey and Peng Zhao}
\affiliation[]{Department of Applied Mathematics and Theoretical Physics, \\University of Cambridge, United Kingdom}
\emailAdd{N.Dorey@damtp.cam.ac.uk}
\emailAdd{P.Zhao@damtp.cam.ac.uk}

\abstract
{We study scalar excitations of high spin operators in $\mathcal{N}=4$
  super Yang-Mills theory, which are dual to solitons propagating on a long
  folded string in $AdS_{3} \times S^{1}$. In the spin chain
  description of the gauge theory, these are associated to
  holes in the magnon distribution in the 
 $\mathfrak{sl}(2,\mathbb{R})$ sector. We compute the
  all-loop hole S-matrix from the asymptotic Bethe ansatz, and expand
  in leading orders at weak and strong coupling. The worldsheet
  S-matrix of solitonic excitations on the GKP string is calculated
  using semiclassical quantization. We find an exact agreement between
  the gauge theory and string theory results.}

\notoc
\begin{document}
\begin{flushright} DAMTP-2011-33 \end{flushright}
\maketitle

\newpage

\section{Introduction}
The discovery of integrability on both sides of the AdS/CFT
correspondence has opened the possibility for precise tests of the
conjecture. On the gauge theory side, the full spectrum of planar
$\mathcal{N}=4$ super Yang-Mills theory has been determined by the
all-loop asymptotic Bethe ansatz \cite{Beisert:2005fw} for an
integrable $\mathfrak{psu}(2,2|4)$ spin chain. Of particular interest
is the $\mathfrak{sl}(2)$ subsector of twist operators composed of
light-cone covariant derivatives $D_{+}$ acting on 
complex scalar fields $Z$, 
\eq{
\text{tr}(ZD_{+}^S Z)+\cdots,
\label{sl2operator}
} 
where the ellipsis denotes all other operators that mix under
renormalization. At large Lorentz spin $S$, 
the conformal dimension $\Delta$ of the ground state in this sector 
exhibits logarithmic scaling at all loops
\eq{
E = \Delta - S = 2\Gamma_{\text{cusp}}(g)\log S + \mathcal{O}(\log^{0} S).
} 
The cusp anomalous dimension $\Gamma_\text{cusp}$ interpolates
smoothly between small and large values of the coupling constant $g =
\sqrt{\lambda}/4\pi$ \cite{Basso:2007wd}.\footnote{For a review of twist
  operators and the cusp anomalous dimension, see
  \cite{Freyhult:2010kc}.} It also controls the infrared divergences
of gluon scattering amplitudes, which are dual to light-like polygonal
Wilson loops. There has been much recent interest in studying
excitations around this ground state \cite{Alday:2007mf, Gaiotto:2010fk,
  Giombi:2010bj}. In particular, their spectrum is an important ingredient
in the operator product expansion of light-like Wilson loops \cite{Alday:2010ku}. 
The spectrum of all such excitations has been determined by 
Basso \cite{Basso:2010in}, by way of a set of integral equations for
the excitation densities which can be solved at all values of the
coupling. The scalar excitations on top of the sea of derivatives
(\ref{sl2operator}) are known as holes. 
In the spin chain language, the holes are gaps in the magnon distribution.
The vacuum contains two large holes of order $\mathcal{O}(S)$ 
which contribute to the logarithmic scaling \cite{Belitsky:2006en}. Excitation of the 
vacuum introduces additional holes of order $\mathcal{O}(S^{0})$.

The string state dual to the ground state operator of fixed spin
(\ref{sl2operator}) 
is a long folded string rotating in $AdS_{3} \times S^1$, known as the
GKP string \cite{Gubser:2002tv}. The string energy also exhibits
logarithmic scaling due to the length of the string in the target
space, whose coefficient coincides with $\Gamma_\text{cusp}$ at strong
coupling. The two spikes at the end of the string approach the
boundary of $AdS_{3}$ and are identified with the two large
holes. More generally 
hole excitations correspond to classical solitons that propagate on
the GKP string. These are identified with solitons of the sinh-Gordon
equation after Pohlmeyer reduction of the string equations of motion 
\cite{Jev}. The classical dispersion relation for these excitations was
obtained in \cite{Dorey:2010iy} where they 
were named ``Giant Holes'' in analogy with the Giant Magnons of 
\cite{Hofman:2006xt}. 

In this paper we study the scattering of two holes/solitons for all
values of the coupling. On the gauge
theory side, the scattering phase is obtained by considering the
scattering of the second hole with the change in magnon density due to
the first hole \cite{Faddeev:1977rm}. Following the techniques
introduced in \cite{Basso:2010in, Basso:2009gh}, we obtain an integral
expression, which we expand at weak and strong coupling. On the string
theory side, the worldsheet S-matrix is calculated from the time-delay
\cite{Jackiw:1975im} that one soliton accumulates as it passes the
other. It receives additional contributions due to the non-trivial 
relation between worldsheet coordinates and global coordinates and
also due to the contribution of each excitation to the length of the 
string. Once these subtleties are carefully taken into account, 
we find exact agreement with the gauge theory 
result at strong coupling.

\section{Scattering of spin chain holes}
First it is important to be precise about we mean by hole excitations.
Consider operators of arbitrary twist 
\eq{
\text{tr}(D_{+}^S Z^{J})+\cdots. 
\label{twistJ}
}
In the spin chain description, we identify the 
$S$ covariant derivatives with magnons propagating on a 
background of
scalars corresponding to the ferromagnetic ground state. 
An important feature of the non-compact
$\mathfrak{sl}(2,\mathbb{R})$ spin chain is that there can be an
arbitrary number of magnons at each spin site without increasing the
length of the chain which is identified 
with the number $J$ of scalars. In addition, the number of holes in the mode
number distribution of the magnons is also equal to $J$. 
The Bethe ansatz equation provides a quantization
condition for the magnon rapidities. The large spin limit we will
consider represents a
highly excited state in which the total length is held fixed. In this
limit, it is natural to view the sea of derivatives as the 
pseudovacuum on which
the scalars act as hole excitations. Moreover, Alday and Maldacena
have identified the elementary fluctuations around the GKP string with
insertion of twist-one operators in the pseudovacuum
\cite{Alday:2007mf}. With these motivations, we define the holes as
vacancies in the magnon distribution, to be identified with inserting
a scalar $Z$, as opposed to removing a derivative $D_{+}$ 
in the trace (\ref{twistJ}). 
\subsection{One-loop S-matrix at weak coupling}
The one-loop Bethe ansatz equation
for $S$ magnon excitations in an 
$\mathfrak{sl}(2,\mathbb{R})$ spin chain of $J$ sites is
\eq{
1 = e^{-ip(u_{k})J}\prod_{\ell \ne k}^{S} S(u_{k},u_\ell),
\label{oneloop}
} 
where the magnon momentum and S-matrix are defined as
\eq{
p(u_{k}) = i\log \frac{u_{k}-i/2}{u_{k}+i/2}, \qquad S(u_{k}, u_{\ell}) = \frac{u_{k} - u_\ell - i}{u_{k} - u_\ell + i}. 
}
Each Bethe root $u_{k}$ is associated with a mode number $n(u_{k})$ corresponding to a branch of the logarithm 
\eq{
0 = 2\pi i n(u_{k}) - 2\sum_{\ell\ne k}^{S}\log\frac{u_{k} - u_\ell - i}{u_{k} - u_\ell + i}  -J\log\frac{u_{k} - i/2}{u_{k} + i/2} .
} 
The roots all lie on the real axis. When $J=2$, the mode numbers are symmetrically distributed as $n(u_{k}) = k, \quad k = \pm\frac{1}{2}, \pm\frac{3}{2}\ldots, \pm \frac{S-1}{2}$.\footnote{For arbitrary $J$, the magnon mode number develops a gap in the middle corresponding to $J-2$ small holes. The gap closes in the large spin limit.} 
In the large spin limit, the magnon roots form a cut in the interval $\left[-\frac{S}{2}, \frac{S}{2}\right]$ and
(\ref{oneloop}) becomes an integral equation for the magnon density $\rho(u) = \frac{dn}{du}$ \cite{Eden:2006rx}
\eq{
0 = \pi \rho(u) - \int_{-\frac{S}{2}}^{\frac{S}{2}} dv~\frac{\rho(v)}{1+(u-v)^2} - \frac{4}{1+4u^2}.
}
Introducing an excited hole of rapidity $u_{h}$ shifts $J=2 \mapsto 3$ and the magnon mode number $n(u_{k}) \mapsto n(u_{k}) + \theta(u - u_{h})$, where $\theta(u)$ is the Heaviside step function. The change in the ground state density $\rho_{0}(u)$ due to the hole, $\rho_{h}(u, u_{h})$, can be written as
\eq{
\rho_{0}(u) + \rho_{h}(u,u_{h}) = \rho(u) + \delta(u-u_{h}).
\label{excitation}
}
We will also use $\rho_{h}(u)$ when we do not need to specify the hole rapidity. It satisfies the integral equation 
\eq{
0 = \pi\rho_h(u) - \int dv~\frac{\rho_h(v)}{1+(u-v)^2} + \frac{1}{1+(u-u_{h})^2} - \frac{2}{1+4u^{2}} .
}
Note that we have extended the integration domain to the entire real line. This is justified because the excitations are of order $\mathcal{O}(S^{0})$ and $\rho_{h}(u)$ is suppressed at large rapidity. We solve for $\rho_h(u)$ by Fourier transform to obtain 
\eq{
\rho_{h}(u) = \psi\left(\frac{1}{2} + i u\right)+\psi\left(\frac{1}{2} - i u\right) - \psi\left(1 + i (u-u_{h})\right) - \psi\left(1 - i(u-u_{h})\right),
\label{density}
}
which decays as $\rho_{h}(u) = \frac{2u_{h}}{u} + \mathcal{O}(u^{-2})$ at infinity.
 
The S-matrix between two holes carrying rapidities $u_{h,1}, u_{h,2}$ is a phase $S(u_{h,1},  u_{h,2}) = -\exp (i\delta_{\rm spin})$. It is given by  integrating the scattering of the second hole with the magnon density change introduced by the first hole \cite{Faddeev:1977rm}.
\eq{ 
\partial_{u_{h,2}}\log S(u_{h,1}, u_{h,2}) = \int du~ \partial_{u}\log S(u,u_{h,2})\left(\rho_h(u, u_{h,1}) -\delta(u-u_{h,1})\right)\label{phase}.
}
The integral can be evaluated by Fourier transform. We find
\EQ{
\delta_{\rm spin} = &-i\log \frac{\Gamma(1-i(u_{h,1} - u_{h,2}))}{\Gamma(1+i(u_{h,1} - u_{h,2}))}-i\log\frac{\Gamma(1/2+iu_{h,1})}{\Gamma(1/2-iu_{h,1})}
-i\log\frac{\Gamma(1/2-iu_{h,2})}{\Gamma(1/2+iu_{h,2})} \\ & - p(u_{h,1}) + p(u_{h,2}) ,\label{oneloopphase}
}
where in recovering the phase $\delta_{\rm spin}$ from its derivative (\ref{phase}) we fixed the constant of integration such that the expression is antisymmetric in $u_{h,1}$ and $u_{h,2}$, as is necessary for a unitary S-matrix. Note the appearance of momentum terms $p(u_{h,1})$ and $p(u_{h,2})$. They are present because by introducing a hole we have created a vacancy in the magnon distribution. In defining the effective Bethe ansatz for holes, it is more appropriate to absorb such terms as part of the phase due to propagation instead of as part of the S-matrix. \footnote{We thank Benjamin Basso for insightful comments on interpreting the S-matrix and its connection to the effective Bethe ansatz.} Assuming factorized scattering, we may write down the effective Bethe ansatz for a twist $J$ chain
\eq{
1 = \prod_{j\ne i}^{J} S(u_{h,i}, u_{h,j}).
}
The two large holes $u_{h,1}, u_{h,J}$ of order $\mathcal{O}(S)$ are non-dynamical because their positions are fixed by the quantum number of the spin chain. 
We separate their contribution from the other dynamical holes. Using Stirling's approximation, we find  
\eq{
1 = e^{-ip_{h}(u_{h,i})\times 2\log S + \mathcal{O}(S^{0})} \prod_{j=2}^{J-1} \til S(u_{h,i}, u_{h,j}), \label{effective}
}
where $p_{h}(u_{h}) = 2u_{h} + \mathcal{O}(g^{2})$ is the hole momentum \cite{Belitsky:2006en} and we redefine the S-matrix as
\eq{
\til S(u_{h,i}, u_{h,j}) = -\frac{\Gamma(1-i(u_{h,i} - u_{h,j}))}{\Gamma(1+i(u_{h,i} - u_{h,j}))}\frac{\Gamma(1/2+iu_{h,i})}{\Gamma(1/2-iu_{h,i})}\frac{\Gamma(1/2-iu_{h,j})}{\Gamma(1/2+iu_{h,j})}. 
\label{S-matrix}
}
The two large holes define the effective length of the chain (=$2\log S$) as seen by the other dynamical holes. There are $\mathcal{O}(S^{0})$ contribution to the effective length from the momentum terms in (\ref{oneloopphase}) that was absorbed into the propagation phase, as well as an additional phase $\displaystyle -2i\log \frac{\Gamma(1/2+iu_{h,i})}{\Gamma(1/2-iu_{h,i})}$ from separating the two large holes. They are suppressed at large $S$.
The effective Bethe ansatz (\ref{effective}) is equivalent to the quantization conditions of \cite{Belitsky:2006en} (see equation (3.41) of this reference). 

By comparing (\ref{density}) and (\ref{S-matrix}), we note that the excitation density $\rho_{h}(u,u_{h})$ coincides with the hole scattering kernel $i\partial_{u_{h}}\log \til S(u,u_{h})$. This is not surprising because the definition of $\rho_{h}(u)$ (\ref{excitation}) can be rewritten as an integral equation for the hole density $\delta(u-u_{h})$
\eq{
\rho_{0}(u) - \rho(u)  =  \delta(u-u_{h}) - \int dv~ \rho_{h}(u,v) \delta(v-u_{h}).
}
Hence $\rho_{h}(u)$ can be naturally interpreted as the hole scattering kernel. This general fact has also been observed in the study of Destri-de Vega type non-linear integral equations \cite{Fioravanti:1996, Freyhult:2007pz}.

\subsection{All-loop asymptotic Bethe ansatz}
The all-loop generalization of (\ref{oneloop}) is given by the asymptotic Bethe ansatz equation \cite{Beisert:2005fw}
\eq{
1 = \left(\frac{x^{-}_{k}}{x^{+}_{k}}\right)^{J}\prod_{\ell \ne k}^{S} 
S(u_{k},u_\ell)e^{2i\theta(u_{k},u_\ell)}, 
}
where the magnon S-matrix is defined by the deformed rapidity $u_{k} \pm \frac{i}{2}= x_{k}^{\pm} + \frac{g^2}{x_{k}^\pm}$ as 
\eq{
S(u_{k}, u_\ell) = \frac{x^{-}_{k}-x^{+}_{\ell}}{x^{+}_{k}-x^{-}_{\ell}}\left(\frac{1-g^{2}/(x_{k}^{+}x_{\ell}^{-})}{1-g^{2}/(x_{k}^{-}x_{\ell}^{+})}\right),
}
and $\theta(u,v)$ is the BES dressing phase \cite{Beisert:2006ez} that gives the correct answer beyond three loops in weak-coupling and at leading order in strong coupling. As in the one-loop case, we obtain an integral equation for the excitation density $\rho_{h}(u)$ in the large spin limit
\EQ{
0 &= 2\pi\rho_{h}(u) + \int dv \left[i\partial_{u}\log S(u, v) - 2\partial_{u}\theta(u,v)\right]\rho_{h}(v)\\
& \quad + 2\partial_{u}\theta(u,u_{h}) - i \partial_{u}\log S(u,u_{h}) + i\partial_{u}\log \left(\frac{x^{-}_{k}}{x^{+}_{k}}\right).
}
Following \cite{Basso:2010in}, we split the density into parity even and odd parts in $u$ as
$\rho_{h}(u) = -\sigma_{h}(u) - \til\sigma_{h}(u)$. We also split the magnon scattering kernel $i\partial_{u}\log S(u, v)$ into $\mathcal{K}^{\circ}, \til{\mathcal{K}}^{\circ}$, the dressing contributions into $\mathcal{I}_{\rm dressing}, \til{\mathcal{I}}_{\rm dressing}$, and the inhomogeneous term $i \partial_{u}\log S(u,u_{h})$ into $\mathcal{I}_{h}, \til{\mathcal{I}}_{h}$. 
From their integral expressions \cite{Basso:2010in}, Basso observed that the main scattering kernels are also even and odd in $v$. Thus the parity even and odd parts decouple and we obtain an integral equation for each
\EQ{
0 &= 2\pi \sigma_{h}(u) + \int dv~\mathcal{K}^{\circ}(u,v)~\!\sigma_{h}(v) + \mathcal{I}_{\rm dressing}(u) + \mathcal{I}_{h}(u), \\
0 &= 2\pi \til\sigma_{h}(u) + \int dv~\til{\mathcal{K}}^{\circ}(u,v)~\!\til\sigma_{h}(v) + \til{\mathcal{I}}_{\rm dressing}(u) + \til{\mathcal{I}}_{h}(u).\label{integral1}
}
We proceed to solve for the densities and compute the S-matrix between two excited holes. As before, the scattering phase is given by integrating the excitation density against the inhomogeneous term
\EQ{
\partial_{ u_{h,2}} \delta_{\rm spin} = &\int d u\left[\mathcal{I}_{h} (u,  u_{h,2})~\sigma_h( u,  u_{h,1}) + \til{\mathcal{I}}_{h} (u,  u_{h,2})~\til\sigma_h( u,  u_{h,1})\right] \\
&- \mathcal{I}_{h} (u_{h,1},  u_{h,2}) - \til{\mathcal{I}}_{h} (u_{h,1},  u_{h,2})
, \label{induced}
}
where the even and odd inhomogeneous terms $\mathcal{I}_{h}, \til{\mathcal{I}}_{h}$ admit the integral representations \cite{Basso:2010in}
\EQ{
\mathcal{I}_{h}( u, u_{h}) & = -2 \int_{0}^{\infty} dt~\cos( u t) e^{-t/2} \pi_{h}(t,  u_{h}), \\  
\til{\mathcal{I}}_{h}( u, u_{h}) &= -2 \int_{0}^{\infty} dt~\sin( u t) e^{-t/2} \til\pi_{h}(t,  u_{h}),\\ 
\pi_{h}(t,  u_{h}) &= \cos( u_h t) e^{-t/2} - J_{0}(2gt) - 4g^{2}\int_{0}^{\infty} ds~tK(2gt, 2gs)\cos( u_h s)e^{-s/2}, \\
\til\pi_{h}(t,  u_{h}) &= \sin( u_h t) e^{-t/2} -  4g^{2}\int_{0}^{\infty} ds~tK(2gt, 2gs)\sin(u_h s)e^{-s/2}.
}
The symmetric BES kernel $K(t,s)$ can be split into parity even and odd parts $K = K_{+}+K_{-}$ that admit expansion over Bessel functions as
\eq{ 
K_{+}(t,s)= \frac{2}{ts}\sum_{n=1}^{\infty}(2n-1)J_{2n-1}(t)J_{2n-1}(s),\qquad  K_{-}(t,s) = \frac{2}{ts}\sum_{n=1}^{\infty}(2n)J_{2n}(t)J_{2n}(s).
}
Define the Fourier-Laplace transformed densities as
\eq{ 
\Omega_{h}(t) = ~e^{-t/2}\!\int d u~ \sigma_{h}( u)\cos(ut), \qquad \til \Omega_{h}(t) = ~e^{-t/2}\!\int d u~ \til \sigma_{h}(u)\sin(ut), 
}
for $t>0$. By using the integral representations for the main scattering kernel, the dressing contributions, and the inhomogeneous terms \cite{Basso:2010in}, one can write the integral equations (\ref{integral1}) as equations for $\Omega_{h}, \til \Omega_{h}$
\EQ{
(e^{t}-1)\Omega_{h}(t) + 4g^{2}\int_{0}^{\infty}ds~t \left(K(2gt, 2gs)\Omega_{h}(s) + K_{-}(2gt,2gs)\frac{q^{h}_{-}(2gs)}{e^{s}-1}\right) &= \pi_{h}(t), \\
(e^{t}-1)\til\Omega_{h}(t) + 4g^{2}\int_{0}^{\infty}ds~t \left(K(2gt, 2gs)\til\Omega_{h}(s) + K_{+}(2gt,2gs)\frac{\til q^{\!\!\ h}_{+}(2gs)}{e^{s}-1}\right) &= \til\pi_{h}(t), \label{integral2}
}
where $q^{h}_{-}, \til q^{\!\!\ h}_{+}$ are the generating functions for the conserved charges sourced by the holes. Their dependence on the excitation densities can be seen from the following integral representations \cite{Basso:2010in}
\EQ{
q_{-}^{h}(2gt) &= 8g^{2}\int_{0}^{\infty}ds~t K_{+}(2gt,2gs)\left(\Omega_{h}(s)+\cos(u_{h}s) e^{-s/2}\right), \\
\til q^{\!\!\ h}_{+}(2gt) &= 8g^{2}\int_{0}^{\infty}ds~t K_{-}(2gt,2gs)\left(\til\Omega_{h}(s)+\sin(u_{h}s) e^{-s/2}\right).
}
One recognizes (\ref{integral2}) as a generalization of the celebrated BES equation \cite{Beisert:2006ez}, which gives an all-loop expression for the density that can be expanded at weak and strong limits of the coupling and is amenable to numerical study at intermediate values of the coupling.
The expression for the scattering phase (\ref{induced}) can be written in terms of $\Omega_{h}, \til \Omega_{h}$ as
\EQ{
\partial_{ u_{h,2}} \delta_{\rm spin} &= -2 \int_{0}^{\infty} dt \Big[\left(\Omega_{h}(t, u_{h,1})+e^{-t/2}\cos(u_{h,1}t)\right) \pi_{h}(t,  u_{h,2}) \\
&\qquad \qquad \qquad + \left.\left(\til\Omega_{h}(t,  u_{h,1})+e^{-t/2}\sin(u_{h,1}t) \right) \til\pi_{h}(t, u_{h,2})\right].
\label{induced2}
}
At weak coupling, we can solve (\ref{integral2}) iteratively by expanding the Bessel function near the origin. We recover the one-loop scattering phase (\ref{oneloopphase}) from (\ref{induced2}).

\subsection{S-matrix at strong coupling}
To expand (\ref{induced2}) at strong coupling, it is useful to rescale $u = 2g~\!\! \conj u$ and introduce the auxiliary densities $\gamma^{h}_{\pm}(t), \til\gamma^{h}_{\pm}(t)$ 
related to $\Omega_{h}(t), \til \Omega_{h}(t)$ by
\EQ{
(e^{t/2g}-1)\Omega_h(t) &= \cos(\conj u_h t)e^{-t/4g} - J_0(t) + \gamma^h_-(t) + \gamma^h_+(t), \\
(e^{t/2g}-1) \til\Omega_h(t) &= \sin(\conj u_h t)e^{-t/4g} + \til\gamma^h_-(t) + \til\gamma^h_+(t).
}
The auxiliary functions $\gamma^{h}_{\pm}, \til\gamma^{h}_{\pm}$ admit expansion as Neumann series of Bessel functions
\eq{
\gamma_{-}^{h} = 2\sum_{n=1}^{\infty} (2n-1) \gamma^{h}_{2n-1}J_{2n-1}(t), \qquad \gamma_{+}^{h} = 2\sum_{n=1}^{\infty} (2n) \gamma^{h}_{2n}J_{2n}(t),
}
and similarly for $\til\gamma_{\pm}^{h}$. We expand (\ref{integral2}) in series of Bessel functions and separate into parity even and odd parts. We obtain four equations, two of which relate the higher conserved charges to the auxiliary functions, $q^{h}_{-} = -2\gamma^{h}_{-}, \quad \til q^{\!\!\ h}_{+} = -2\til \gamma^{\!\!\ h}_{+}$, which can be used to express the other pair solely in terms of the auxiliary functions
\EQ{
\gamma_{n}^{h} + \int_{0}^{\infty} \frac{dt}{t} J_{n}(t)\frac{\gamma^{h}_{+}(t) - (-1)^{n}\gamma^{h}_{-}(t)}{e^{t/2g}-1} &= -\int_{0}^{\infty} \frac{dt}{t} \frac{J_{n}(t)}{e^{t/2g}-1}\left(\cos(\conj u_h t)e^{t/4g} - J_{0}(t)\right), \\ 
\til\gamma^{h}_{n} + \int_{0}^{\infty} \frac{dt}{t} J_{n}(t)\frac{\til\gamma^{h}_{-}(t)+(-1)^{n}\til\gamma^{h}_{+}(t)}{e^{t/2g}-1} &= -\int_{0}^{\infty} \frac{dt}{t} \frac{J_{n}(t)}{e^{t/2g}-1}\sin(\conj u_h t)e^{t/4g}.
\label{integral3}
}
We follow the technique in \cite{Basso:2009gh} to transform the set of equations (\ref{integral3}) into a singular integral equation. Define $\Gamma(t) = \Gamma^{h}_{+}(t) + i\Gamma^{h}_{-}(t)$ related to $\gamma^{h}(t) = \gamma^{h}_{+}(t) + i\gamma^{h}_{-}(t)$ via
\eq{
\Gamma^h_\pm(t)=\gamma_\pm^{h}(t)\mp\coth(t/4g)\gamma_\mp^h(t), \qquad
\til\Gamma^h_\pm(t)= \pm\til\gamma_\mp^{h}(t)+\coth(t/4g)\til\gamma_\pm^h(t).
}
At strong coupling, the first term in the LHS of (\ref{integral3}) are suppressed.
We apply the Jacobi-Anger expansion to replace the infinite system of equations (\ref{integral3}) with a single equation for $\Gamma^h_{\pm}$ and $\til \Gamma^h_{\pm}$
\EQ{
\int_{0}^{\infty} dt~\left[e^{i\conj u t}\Gamma^h_{-}(t) - e^{-i\conj u t}\Gamma^h_{+}(t)\right] &\approx -2\int_{0}^{\infty} dt~\frac{e^{i\conj ut}}{e^{t/2g}-1}\left[\cos(\conj u_{h}t)e^{t/4g} - J_{0}(t)\right],\\
\int_{0}^{\infty} dt~\left[e^{i\conj u t}\til\Gamma^h_{-}(t) - e^{-i\conj u t}\til\Gamma^h_{+}(t)\right] &\approx -2\int_{0}^{\infty} dt~\frac{e^{i\conj ut}}{e^{t/2g}-1}\sin(\conj u_{h}t)e^{t/4g},
}
with  $|\conj u| \le 1$. In the Fourier space, they become singular integral equations for $\hat\Gamma^h(\conj u)$ and $\hat{\til\Gamma}\!\!\ ^h(\conj u)$
\EQ{
\pint dk ~\frac{\hat\Gamma^h(k)}{k-\conj u} + \pi \hat\Gamma^h(\conj u) &
\approx -2g \left[\log \left(4|\conj u^{2}-\conj u_h^{2}|\right)+2\sin^{-1}(\conj u)\right]
\equiv -2\kappa^{h}(\conj u), \\
\pint dk ~\frac{\hat{\til\Gamma}\!\!\ ^h(k)}{k-\conj u} + \pi \hat{\til\Gamma}\!\!\ ^h(\conj u)  
&\approx 2g \left[\pi\text{sgn}(\conj u_h)+\log\left|\frac{\conj u+\conj u_h}{\conj u-\conj u_h}\right|\right]\equiv -2\til\kappa^{h}(\conj u).
}
The above relations hold for $|\conj u| \le 1$. For $|\conj u| > 1$, $\hat\Gamma^{h}(\conj u), \hat{\til\Gamma}\!\!\ ^h(\conj u)$ can be determined by analyticity and is subleading at strong coupling \cite{Basso:2009gh}. Thus we can restrict the range of integration to the unit interval. 
Such singular integral equations have a standard solution
\eq{
\hat \Gamma^h(k) \stackrel{k^{2} \le 1}\approx  -\frac{1}{\pi} \left\{\kappa^{h}(k) - \left(\frac{1+k}{1-k}\right)^{1/4}\left[\frac{1}{\pi}\pint_{-1}^{1}\frac{d\conj u~\kappa^{h}(\conj u)}{\conj u-k}\left(\frac{1-\conj u}{1+ \conj u}\right)^{1/4} - \frac{\sqrt{2}~\!c}{1+k}\right]\right\},
\label{fourier}
}
where $c$ can be determined from the so-called quantization conditions \cite{Basso:2009gh} and is negligible.

The finite Hilbert transform in (\ref{fourier}) can be evaluated using contour integration \cite{King:2009}. As $\kappa^{h}(\conj u),\til \kappa^{h}(\conj u)$ have branch points, it is simpler to consider instead $\partial_{\conj u_{h}}\hat\Gamma^{h}(\conj u)$ and $\partial_{\conj u_{h}}\hat{\til\Gamma}\!\!\ ^h(\conj u)$. We find
\EQ{
&\partial_{\conj u_{h}} \hat\Gamma^{h}(k) \approx 
- \frac{\sqrt{2}~\!g}{\pi} \left(\frac{1+k}{1-k}\right)^{1/4} \left[\left(\frac{\conj u_{h} - 1}{\conj u_{h} + 1}\right)^{1/4}\frac{1}{k-\conj u_{h}} - \left(\frac{\conj u_{h} + 1}{\conj u_{h} - 1}\right)^{1/4}\frac{1}{k+\conj u_{h}}\right],
\\
&\partial_{\conj u_{h}} \hat{\til\Gamma}\!\!\ ^h(k) \approx 
- \frac{\sqrt{2}~\!g}{\pi} \left(\frac{1+k}{1-k}\right)^{1/4}  \left[\left(\frac{\conj u_{h} - 1}{\conj u_{h} + 1}\right)^{1/4}\frac{1}{k-\conj u_{h}} + \left(\frac{\conj u_{h} + 1}{\conj u_{h} - 1}\right)^{1/4}\frac{1}{k+\conj u_{h}}\right].
}
We may now compute the scattering phase (\ref{induced2}) from the transformed densities $\hat \Gamma^h(t), \hat{\til\Gamma}\!\!\ ^{h}(t)$. 
Performing the integral is straightforward but tedious. Remarkably, the answer takes a simple form when expressed in terms of the Zhukovsky variables $v_{i} = \conj u_{h,i} - \sqrt{\conj u_{h,i}^{2} - 1}$, which correspond to the soliton velocities in the string worldsheet
\eq{
\delta_{\rm spin}
=\frac{1}{2g}\left(P_{1}E_{2}-P_{2}E_{1}\right) +2g\left[\left(v_{1} + \frac{1}{v_{1}} - v_{2} - \frac{1}{v_{2}}\right)\log v + \frac{1}{\gamma_{1}\gamma_{2}} \left(\frac{1}{v_{1}} - \frac{1}{v_{2}}\right)\right],
}
where $P_{i}, E_{i}$ are the energy and momentum of the giant holes (\ref{dispersion}), $v$ is the soliton velocity in the center of mass frame (\ref{timedelay2}), and $\gamma_{i} = (1-v_{i}^{2})^{-1/2}$ is the Lorentz factor.
The phase is antisymmetric in $v_{1}$ and $v_{2}$, as is expected from a unitary S-matrix. 

\section{Worldsheet scattering of giant holes}
In this section we review the construction of solitonic excitations on the GKP string \cite{Dorey:2010iy} and study their scattering.
Consider classical strings in $AdS_{3}$, in embedding coordinates
\eq{
-X_0^{2}-X_{1}^{2}+X_{2}^{2}+X_{3}^{2} = -1.
}
We may form two complex coordinates and express in global $AdS_{3}$ 
coordinates $\{\rho,\phi,t\}$ as
\eq{
Z_{1} = X_{0} + iX_{1} = e^{it}\cosh\rho, \qquad Z_{2} = X_{3} + iX_{4}= e^{i\phi}\sinh\rho.
}
Integrability of the classical string is manifest by performing a
Pohlmeyer reduction on the equation of motion to obtain the
sinh-Gordon equation \cite{Jev} 
\eq{
\hat\partial^{2}\hat\alpha + \sinh\hat\alpha = 0.
}
The GKP string is a vacuum solution with sinh-Gordon angle $\hat\alpha = 0$, which yields
\eq{
Z_{1} = e^{i\tau}\cosh\sigma, \qquad Z_{2} = e^{i\tau}\sinh\sigma.
}
As in \cite{Dorey:2010iy}, to regulate the IR divergence in the 
string energy a cut-off $\pm\Lambda$ on the range of the worldsheet
coordinate $\sigma$ is introduced.  
It is natural to consider solitonic excitations of this
vacuum. However, a one-soliton solution does not correspond to a
physical state of the string due to
the zero total momentum constraint 
$\sum_{i} P_{i} = 0$. Instead, we construct a two-soliton solution 
and then identify one of the solitons when it is 
well-separated from the other as the ``one-soliton solution''. 
The two-soliton solution in the center of mass frame is \cite{Jev}
\EQ{
Z_{1}^{ss}(\tau, \sigma) &= e^{i\tau}\frac{v\cosh T\cosh\sigma+\cosh X\cosh\sigma-\sqrt{1-v^{2}}\sinh X\sinh\sigma-i\sqrt{1-v^{2}}\sinh T\cosh\sigma}{\cosh T+v\cosh X}, \\
Z_{2}^{ss}(\tau, \sigma) &= e^{i\tau}\frac{v\cosh T\sinh\sigma+\cosh X\sinh\sigma-\sqrt{1-v^{2}}\sinh X\cosh\sigma-i\sqrt{1-v^{2}}\sinh T\sinh\sigma}{\cosh T+v\cosh X},
\label{twosoliton}
}
where $X = 2\gamma\sigma, T = 2v\gamma\tau$ with sinh-Gordon angle
\eq{
\hat\alpha_{ss}=\log\left(\frac{v\cosh X - \cosh T}{v\cosh X + \cosh T}\right)^{2}.
}
We may study a single soliton by considering a limit of the above
solution where the second soliton is sent to infinity. The solution is
obtained by taking the large $T, X$ limit of (\ref{twosoliton}) and
takes the 
form \cite{Dorey:2010iy}
\EQ{
Z_{1}^{s}(\tau, \sigma) &= e^{i\tau}\frac{e^{\Sigma}(\cosh\sigma-\sqrt{1-v^{2}}\sinh\sigma) + e^{-\Sigma}(v-i\sqrt{1-v^{2}})\cosh\sigma}{ve^{\Sigma}+e^{-\Sigma}}, \\
Z_{2}^{s}(\tau, \sigma) &= e^{i\tau}\frac{e^{\Sigma}(\sinh\sigma-\sqrt{1-v^{2}}\cosh\sigma) + e^{-\Sigma}(v-i\sqrt{1-v^{2}})\sinh\sigma}{ve^{\Sigma}+e^{-\Sigma}},
}
where $\Sigma = (X-T)/2 = \gamma(\sigma - v\tau)$. The sinh-Gordon angle reduces to that of a single soliton so it can be interpreted as a one-soliton solution. 
At large distance the soliton solution can be matched onto the vacuum
or GKP solution using the asymptotics, 
\EQ{
Z_{1}^{s}(\tau, \sigma) &\simeq \frac{1}{2} e^{i\tau}e^{\sigma}\left(\frac{1-\sqrt{1-v^{2}}}{v}\right) \quad \text{as } \sigma \to \infty, \\
Z_{1}^{s}(\tau, \sigma) &\simeq \frac{1}{2} e^{i\tau}e^{-\sigma}\left(v-i\sqrt{1-v^{2}}\right) \quad \text{as } \sigma \to -\infty.
}
In order to match the asymptotics of the GKP string solution, we need
to restrict the worldsheet coordinate to $-\Lambda < \sigma < \Lambda
- \alpha(v)$ where $\Lambda$ is the IR cut-off \cite{Dorey:2010iy}. 
Note also that global AdS time no longer coincides with the
worldsheet time and is shifted by a constant 
$t \simeq \tau - \beta(v)$ as $\sigma \to -\infty$ where we define, 
\eq{
\alpha(v) = \log\left(\frac{1-\sqrt{1-v^{2}}}{v}\right), \qquad
\beta(v) 
= \tan^{-1}\left(\frac{\sqrt{1-v^{2}}}{v}\right).
}
The dispersion relation of the giant holes has been computed in 
\cite{Dorey:2010iy} as,  
\eq{
E(v) = g\left[\log\left(\frac{\gamma+1}{\gamma-1}\right) - \frac{2}{\gamma}\right], \qquad P(v) = 2g\left[\frac{1}{\gamma v} - \tan^{-1}\left(\frac{1}{\gamma v}\right)\right]. \label{dispersion}
}

To move to a frame where the two solitons have generic velocities $v_{1}, v_{2}$, we introduce the following boost 
\EQ{
\begin{pmatrix} \tau \\ \sigma \end{pmatrix} = \begin{pmatrix} \cosh\frac{\theta_{1}+\theta_{2}}{2} & -\sinh\frac{\theta_{1}+\theta_{2}}{2} \\ -\sinh\frac{\theta_{1}+\theta_{2}}{2} & \cosh\frac{\theta_{1}+\theta_{2}}{2}\end{pmatrix} \begin{pmatrix} \tau' \\ \sigma' \end{pmatrix}, 
}
where $\theta_{i}$ is the soliton rapidity. One may easily check that
\eq{
\frac{1}{2} (X - T) = \gamma_{1}(\sigma' - v_{1}\tau'), \qquad
\frac{1}{2} (X + T) = \gamma_{2}(\sigma' - v_{2}\tau').
}
As $\tau' \to \infty$, the two solitons are now located at $\sigma' = v_{i}\tau'$. We may separate them by zooming into the first soliton in the limit $T \to \pm \infty$ and $X \to \pm \infty$ (for the second soliton we would zoom in the limit $T \to \pm \infty$ and $X \to \mp \infty$). The sinh-Gordon angle will be that of a single soliton with a time-shift.
\eq{\hat\alpha_{ss} \to \log\tanh^{2}\left\{\gamma_{1}\left[\sigma' - v_{1}\left(\tau'\mp\frac{\log v}{2\gamma_{1}v_{1}}\right)\right]\right\} \qquad \text{as } \tau' \to \pm \infty. \label{timedelay}
}
We may read off the time delay experienced by the first soliton from (\ref{timedelay}) as
\eq{
\Delta T_{12} = \frac{\log v}{\gamma_{1}v_{1}}, \qquad v^{2} = \frac{\gamma_{1}\gamma_{2}(1-v_{1}v_{2})-1}{\gamma_{1}\gamma_{2}(1-v_{1}v_{2})+1}. \label{timedelay2}
}
The scattering phase $\exp{(i\delta)}$ can be calculated from the semiclassical formula  $\frac{\partial \delta}{\partial E_{1}} = \Delta T_{12}$ \cite{Jackiw:1975im}, where $E_{1} = E(v_{1})$ is the soliton energy evaluated at $v_{1}$. Although $E(v)$ was computed in the center of mass frame, it is intrinsic to the single soliton so is independent of frame. We integrate the time delay to find
\eq{
\delta =
g\left[-\frac{1}{\gamma_{2}v_{2}}\log\left(\frac{\gamma_{1}+1}{\gamma_{1}-1}\right) +\frac{2}{\gamma_{2}}\tan^{-1}\left(\frac{1}{\gamma_{1}v_{1}}\right) + 2\left(v_{1} + \frac{1}{v_{1}}-v_{2}-\frac{1}{v_{2}}\right)\log v\right],
}
where we used the asymptotic condition $\delta \to 0$ as $v_{1} \to 1$ to fix the integration constant.
However, the S-matrix does not appear unitary: $\delta(v_{1},v_{2})
\ne -\delta(v_{2},v_{1})$. This is familiar from the giant magnon case
\cite{Hofman:2006xt} (see equation (3.33) of this reference), 
where a similar problem arose and
was related to an ambiguity in the length assigned to the excitation. 
In our case, the
second soliton changes the string length by
$-\alpha_{2}=-\alpha(v_{2})$. Because the
energy $E = \Delta - S$ is canonically conjugate to $(t+\phi)/2$, we
also need to include the additional time delay
$\beta_{2}=\beta(v_{2})$ 
from the
second soliton  due to the difference between the global AdS time $t$
and the worldsheet time $\tau$. Taking these effects into account, 
the scattering phase is given as $\delta_{\rm string} = \delta -
P_{1}\alpha_{2} + E_{1}\beta_{2}$. It is then antisymmetric in $v_{1}$ and
$v_{2}$ and exactly matches the gauge theory result at strong coupling, i.e. $
\delta_{\rm string} = \delta_{\rm spin}$.

\section{Discussions}

In this paper we have considered the scattering of two holes on a spin
chain of length two. The generalization to arbitrary number of holes
on a chain of arbitrary length is straightforward. The string theory
picture is that of multiple solitons propagating on a folded, more
generally, spiky string \cite{Dorey:2010id}. Classically we know that
these solitons undergo factorized scattering and, as the theory is
integrable, it is natural to
expect that the S-matrix for any number of holes is exactly factorizable for
all values of the coupling. 
In principle, the two hole S-matrix
obtained in this paper, together with the dispersion relation of 
\cite{Dorey:2010iy}, can therefore 
be used to write down a dual set of asymptotic
Bethe ansatz equations which determine the full spectrum of excited
states of a GKP string of large but finite length 
(in the non-compact $SL(2)$ sector). This must be, of course, equivalent
to the original all-loop asymptotic Bethe ansatz equations of 
\cite{Beisert:2005fw} but should be a simpler starting point for
discussing generic states of large spin.   
\acknowledgments
We thank Benjamin Basso, Diego Bombardelli, Davide Fioravanti, Paolo Grinza, Sungjay Lee, Manuel Losi, Marco Rossi and 
Pedro Vieira for helpful discussions. We also thank Manuel Losi for
making available to us his notes on the semiclassical phase shift. 
PZ is supported by a Dorothy Hodgkin Postgraduate Award from EPSRC and 
Trinity College, Cambridge.

\end{document}